\documentclass{sf2a-conf2011}
\usepackage{graphicx}
\usepackage{hyperref}
\usepackage[]{natbib}  

\def\BibTeX{{\rm B\kern-.05em{\sc i\kern-.025em b}\kern-.08em
    T\kern-.1667em\lower.7ex\hbox{E}\kern-.125emX}}
\bibpunct{(}{)}{;}{a}{}{,}  

\begin{document}

\TitreGlobal{SF2A 2011}


\title{Coincident searches between gravitational waves and high-energy neutrinos with the ANTARES and LIGO/Virgo detectors}

\runningtitle{Coincident searches between GW and HEN}

\author{B. Bouhou, for the ANTARES Collaboration, the LIGO Scientific Collaboration and the Virgo Collaboration}\address{
AstroParticule et Cosmologie (APC) Universit\'e Paris-Diderot, 10 rue Alice Domon et L\'eonie Duquet, 75205 Paris Cedex 13 FRANCE\\
\texttt{bouhou@apc.univ-paris7.fr}}

\setcounter{page}{237}


\maketitle

\begin{abstract}
A multi-messenger approach with gravitational-wave transients and high-energy neutrinos is expected to open new perspectives in the study of the most violent astrophysical processes in the Universe. In particular, gamma-ray bursts are of special interest as they are associated with astrophysical scenarios predicting significant joint emission of gravitational waves and high-energy neutrinos. Several experiments (e.g. ANTARES, IceCube, LIGO and Virgo) are currently recording data and searching for those astrophysical sources. In this report, we present the first joint analysis effort using data from the gravitational-wave detectors LIGO and Virgo, and from the high-energy neutrino detector ANTARES. 
\end{abstract}



\section{Introduction}

With the construction of the ANTARES neutrino telescope (~\cite{Collaboration:2011nsa}) in the Mediterranean Sea and IceCube at the South Pole, together with the interesting sensitivity reached by the LIGO (~\cite{abbott:2004:517}) and the Virgo (~\cite{Virgo}) detectors, the multi-messenger astronomy with gravitational waves (GW) and high-energy neutrinos (HEN) is entering a very promising era. 

ANTARES is an array of 12 strings spaced over an area of 0.1 km$^{2}$, each one holding 75 optical modules. The main goal of the experiment is to search for HEN of astrophysical origin (i.e, with energies $> 100$ GeV) by detecting the Cerenkov photons produced by relativistic muons induced by neutrino charged current interaction in the vicinity of the detector.   
The detector was completed on May 30, 2008 but it started recording data in an incomplete configuration since Jan 30, 2007.

The US project LIGO and European Virgo are Michelson-Morley interferometers, using a very sophisticated technology to measure any variation in the arms length due to the passage of GWs, which are a ripples of space-time caused by accelerating masses. LIGO and Virgo follow the similar operational design but include differences in the arm length for instance (4 km for LIGO and 3 km for Virgo). The interferometer is maintained on the nominal working point where the light beams from each arm interfere destructively, i.e the dark fringe. If a GW passes through the detector it produces a differential strain between the two arms, and hence a change in the phase shift between beams at recombination. This small perturbation results in a fluctuation in the light power after recombination which is converted into the GW strength or strain with $h_{strain} = 2 \Delta L/L$, where $\Delta L$ is the change in separation of two masses at distance $L$. Those detectors have reached their initial sensitivity and are currently being upgraded to improve their sensitivity by one order of magnitude. 

A working group gathering people from the network of experiments ANTARES, IceCube, LIGO and Virgo study the connections between GWs and HENs emitted by astrophysical phenomena such as gamma-ray burst. In Sec \ref{strategy} we discuss the GW and HEN joint search strategy and in Sec \ref{dataanalysis} we describe the analysis of the first combined 2007 data sample from ANTARES (5L), LIGO (S5) and Virgo (VSR1). 

\section{Joint search with ANTARES, LIGO and Virgo data}
\label{strategy}

\subsection{Feasibility and data sets}

\begin{figure}[ht!]
 \centering
\includegraphics[width=0.8\textwidth,clip]{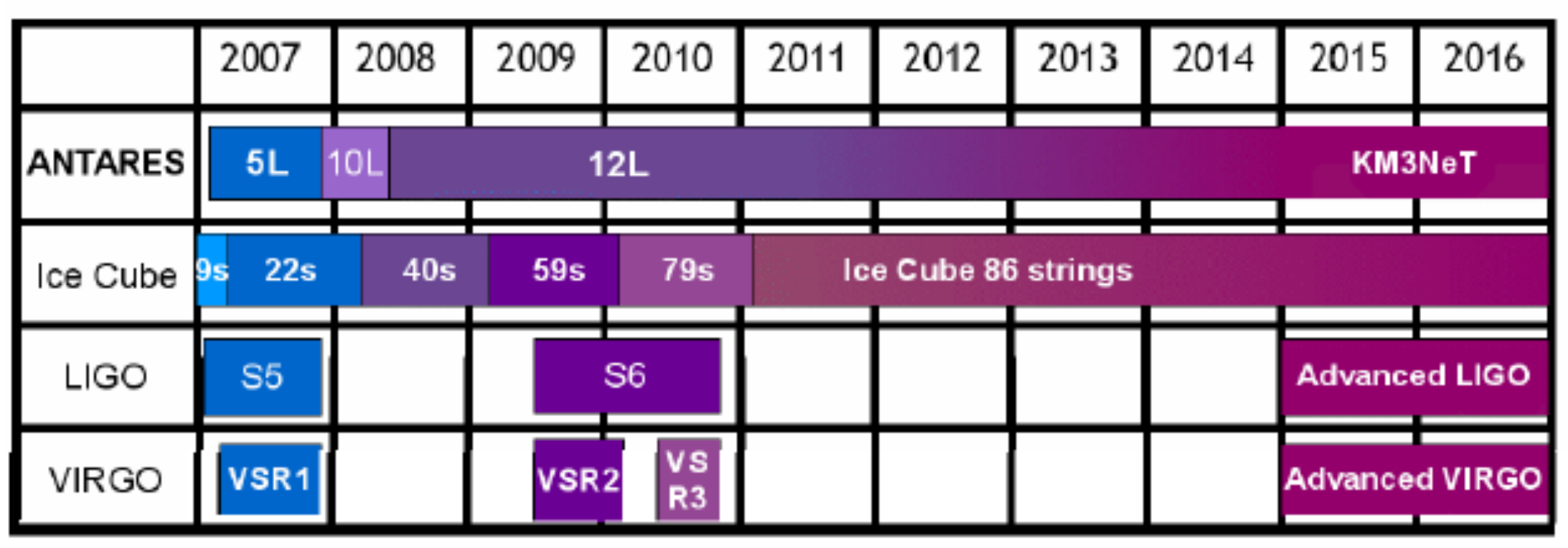}      
  \caption{Time chart of the ANTARES, IceCube, KM3NET, LIGO and the Virgo experiments}
  \label{bouhou:fig1}
\end{figure}

We search for time and spatial coincidences between GW and HEN signals. This is feasible since ANTARES and LIGO/Virgo (~\cite{VanElewyck:2009pf,eric}) share a common view of $\sim$30$\%$ of the sky and several periods of concomitant data takings can be identified.
In late 2007 ANTARES, IceCube, LIGO and Virgo completed a first concomitant data taking period. LIGO completed the fifth "science run", S5, from November 4, 2005 until September 30, 2007. The first Virgo Science Run (VSR1), covered the period from May 18, 2007 until September 30, 2007 (~\cite{Abadie:2010uf}). During this period, ANTARES was operating in 5 active detection strings (5L) and IceCube in 22 active strings. The concomitant set of ANTARES (5L), VSR1 and S5 data, covers the period between January 27th and September 30th, 2007.  A second step will concern the analysis of the sixth LIGO science run, S6, covering the period from July 7, 2009 until December 31, 2010, the second Virgo science run VSR2, covering the period from July, 2009 until December 31 and ANTARES 10L and 12L from the end of December 2007 up to now.
Future schedules involving next-generation detectors with a sensitivity increased by at least one order of magnitude (such as KM3NeT and the Advanced LIGO/Advanced Virgo projects (~\cite{Smith:2009bx})) are likely to coincide as well. A time chart of the experiments is shown in Fig. ~\ref{bouhou:fig1}. Here we report on the first data set taken in 2007.

\subsection{Time and spatial coincidence between gravitational waves and high-energy neutrinos}

\subsubsection{Time search window}
\label{time}

\begin{figure}[ht!]
 \centering
\includegraphics[width=0.8\textwidth,clip]{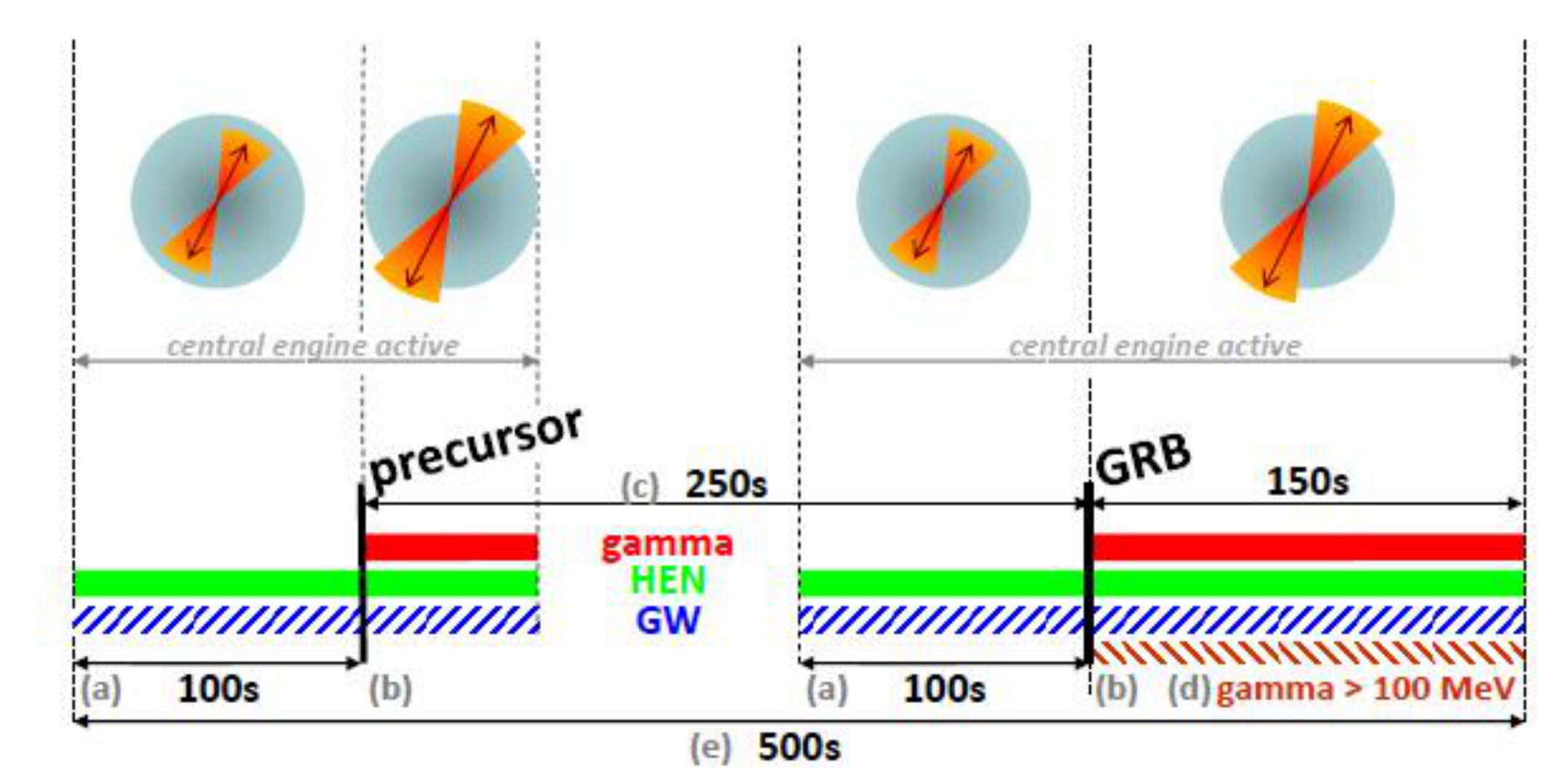}      
  \caption{Overview of the GRB emission processes and GW/HEN time search window. (a) Active central engine before the relativistic jet has broken out of star; (b) Active central engine with relativistic jet broken out of star; (c) Delay between onset of precursor and main burst; (d) Duration corresponding to 90$\%$ of GeV photon emission; (e) Time span of central engine activity. Possible GW/HEN emission between the precursor and the main GRB (no emission is shown on the figure) has no effect on the estimated time window. Overall, the considered processes allow for a maximum of 500s between the observation of a HEN and a GW transient, setting the time search window to [-500s; 500s]. (top) Schematic drawing of plausible emission scenario. For both precursor and main GRB, after the central engine
drives an outflow that breaks out after$\leq$100s. (Adapted from ~\cite{Baret:2011tk})}
  \label{bouhou:fig2}
\end{figure}

Gamma-Ray Bursts (GRBs) are a promising class of extragalactic joint sources of GW and HEN. GRBs are commonly explained by invoking jets of relativistics particles ejected by a yet-to-be-determined ``central engine(s)''. The observed
gamma-rays result from the decay of shock accelerated electrons in the jets. Similarly HENs are expected to be produced by accelerated protons in the same relativistic shocks. The astrophysical systems mentioned as possible central engines are coalescing binaries involving black holes and/or neutron stars or the collapse of massive spinning stars, both expected sources of GW.
GRBs provide an interesting astrophysical scenario where the delay between GW and HEN emissions can be characterized. A conservative estimate of this delay determines the baseline duration over which GW and HEN are declared in coincidence. A statistical estimate has been obtained in (~\cite{Baret:2011tk}). In this article, the authors considered the durations of the different emission processes from GRBs (see Fig.~\ref{bouhou:fig2}), mainly observed by BATSE, Swift and Fermi LAT to infer the size of the time search window. This leads to an upper bound on the size of the time search window. The latter is $\Delta t_{GW+HEN}=[-500s, +500s]$ which is conservative enough to encompass most theoretical models of GW and HEN emissions for GRBs.

\subsubsection{Angular search window}
\label{asw}

\begin{figure}[ht!]
 \centering
 \includegraphics[width=0.48\textwidth,clip]{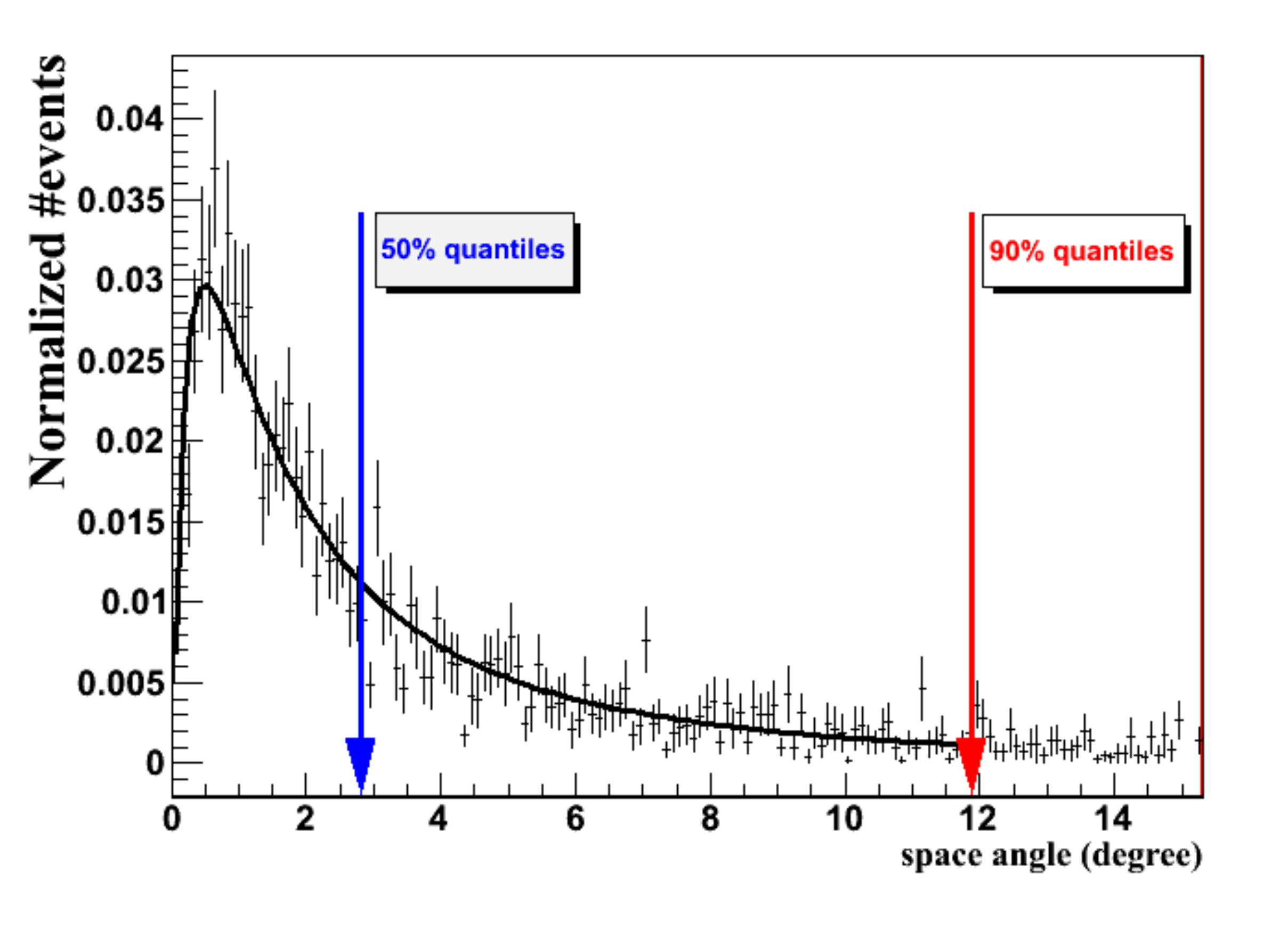}      
 \includegraphics[width=0.50\textwidth,clip]{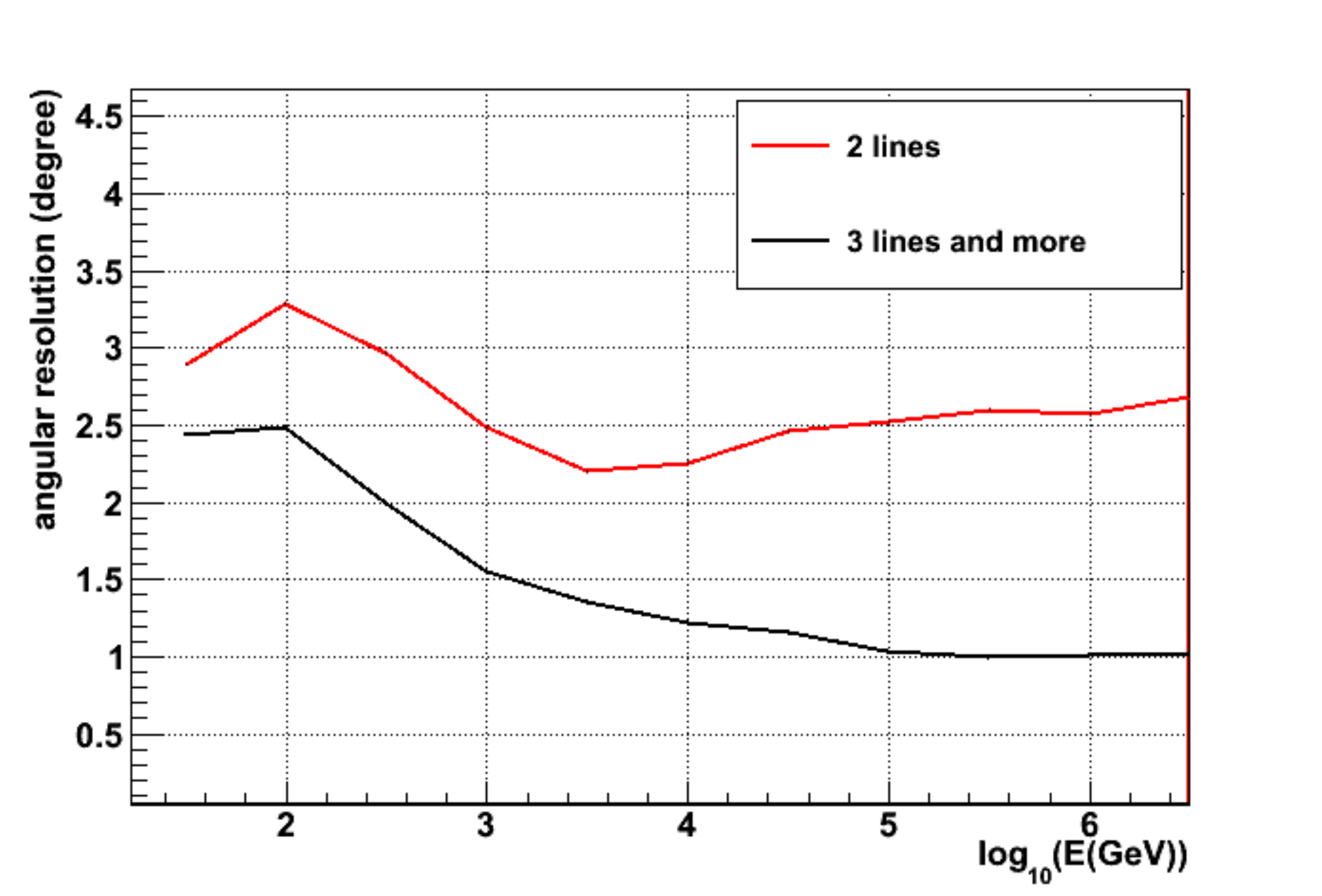}       
 \caption{Left: The error distribution on the neutrino direction for declination between -20 and -10 degrees for event with 7 hits along with its log-normal fit. Right: The angular resolution for ANTARES 5L data represented as function of the true neutrino energy, for events reconstructed with exactly 2 lines (red curve) and for events reconstructed with 3 lines and more (black curve).}
  \label{bouhou:fig3}
\end{figure}

 The angular search window (ASW) is the error distribution of the HEN direction, $\beta$, where 
$\beta$=$|\theta_{true}^{\nu}-\theta_{rec}^{\mu}|$, with $\theta_{true}^{\nu}$ is the true zenith angle of the HEN and $\theta_{rec}^{\mu}$ is the zenith angle of the reconstructed muon track. The angular resolution is usually defined by the median of $\beta$ obtained from a Monte-Carlo simulation sample. Fig. ~\ref{bouhou:fig3} left shows an example of $\beta$ in a bin of declination and Fig. ~\ref{bouhou:fig3} right illustrates the angular resolution as a function of the log-energy. The angular resolution is limited by the detector geometry and by the propagation characteristics of the Cerenkov light in the medium (i.e., photon scattering and absorption). For the cuts defined for this analysis (see Sec.~\ref{selection}), the angular resolution is about one degree for events reconstructed with three lines and more above 100 TeV and of 2.5 degrees at low energies. For events reconstructed with two lines, it ranges from about 2 to 3 degrees.

The radius used for the joint analysis is defined as the 90$\%$ quantile  of $\beta$ (denoted $ASW^{90\%}_{GW+HEN}$). The value of this radius is calculated in bins of the reconstructed declination and in bins of number of hits used in the reconstruction. The distribution of $\beta$ is fitted by a log-normal distribution on an event-by-event basis (see Fig. \ref{bouhou:fig3}).

The coincidence search strategy is as follows: we first select a set of HEN candidates by applying the procedure and cuts detailed in Sec.~\ref{selection} and estimate for each candidate its time of arrival, its ASW and the parameters of the corresponding log-normal fit. These parameters are fed to a search pipeline which tests the presence of a coincident GW signal consistent in time (i.e., within the time coincidence window defined in Sec.~\ref{time}) and direction (i.e., within the angular search window defined in Sec.~\ref{asw}). In practice, we use the so-called X-pipeline  (~\cite{Sutton:2009gi}) algorithm to search coherently the GW data and scan the sky area centered at HEN sky location within a radius $ASW^{90\%}_{GW+HEN}$ (see Sec.~\ref{Xpipeline}). At this point, it is interesting to note that $ASW^{90\%}_{GW+HEN}$ is comparable in size to the typical GW error box (obtained when reconstructing the source direction from the triangulation of GW data).

\section{Data analysis of ANTARES 5L--LIGO/Virgo S5/VSR1 data set}
\label{dataanalysis}

\subsection{Selection of the high-energy neutrino candidates}
\label{selection}

\begin{figure}[ht!]
 \centering
 \includegraphics[width=0.48\textwidth,clip]{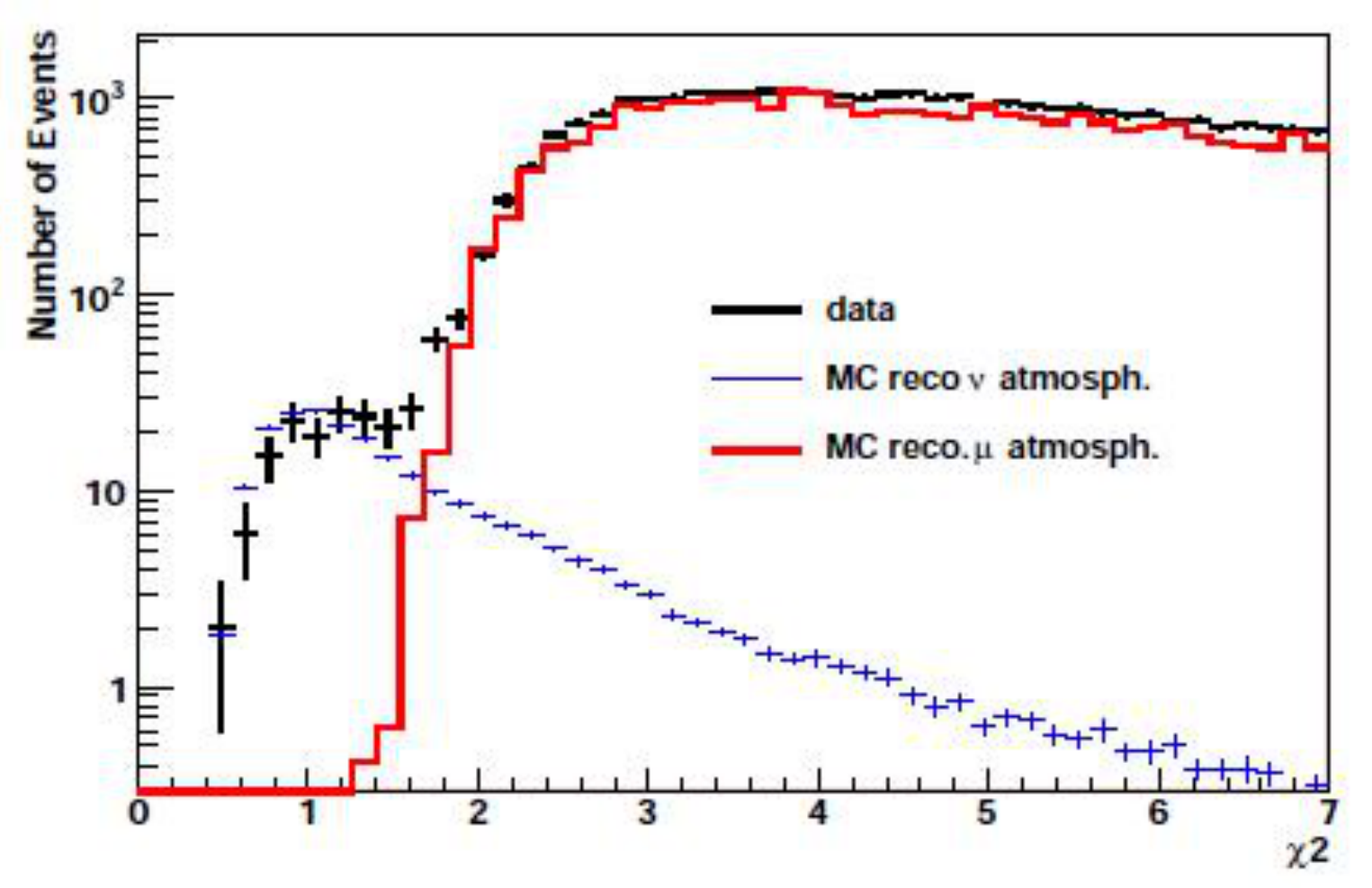}      
  \caption{Distribution of the up-going HEN events as a function of the track quality parameter $\chi^{2}$. The distributions obtained with Monte-Carlo simulations are compared to the data.}
  \label{bouhou:fig4}
\end{figure}

The ANTARES neutrino telescope recorded data from January 1st to December 3rd 2007 with 5 lines. The data sample used in this analysis covers the period between 1st January and 30 September, 2007, as it overlaps with LIGO S5 and Virgo VSR1. These data are sampled using several selection criteria, trigger levels and selection cuts, used to discard the background present at the ANTARES site. Trigger decisions are based on the calculations done at three levels. The first level of trigger is a simple threshold of about 0.5 photo-electron (pe) equivalent charge applied to the analog signal of the photo-multiplier tubes (~\cite{Ageron:2007rj}). The second level trigger is based on coincident hits in the same storey within 20 ns, and hits with large charge (greater than 3 pe or 10 pe depending on the detector configuration). The third level evaluates the characteristics of the hits from the second trigger level. The track reconstruction is based on the $\chi^{2}$-minimization approach implemented in the track reconstruction algorithm (~\cite{Aguilar:2011zz}). Events are identified as sets of hits (direct Cerenkov photons) in a time window of 2.2$\mu$s (~\cite{Aguilar:2010sf}) over the full detector. For the background estimation, various samples of neutrinos and atmospheric muons were simulated and used for data vs. Monte-Carlo comparisons. Various parameters are used to select up-going neutrino candidates and reject physical background (i.e. atmospheric neutrinos and down-going atmospheric muons that are misreconstructed as up-going). Those include the $\chi^2$ of the best fit track, the number of hits used in the fit and the estimated direction of the reconstructed track. For this analysis we define two cut values on the $\chi^2$ depending on the track direction, i.e $\chi^{2}$ $\leq $ 1.8  when $\theta$ $\leq$80$^\circ$ and $\chi^{2}$ $\leq$ 1.4 when 80$^\circ$ $\leq$ $\theta$ $\leq$ 90$^\circ$ where the contamination from down-going muons is higher (see Fig. ~\ref{bouhou:fig4}). The values of the $\chi^{2}$ cuts were optimized based on the maximization of the model discovery potential according to a standard $E^{-2}$ spectrum (~\cite{Becker:2007sv}). With these cuts around $20\%$ of contamination from atmospheric muons remains in the final selected sample. 

After applying the set of selections exposed in the previous section, a sample of 216 neutrino candidates is selected. Each candidate is defined by its arrival time $t_{HEN}$ in the detector, its direction ($\delta_{HEN}$, $\alpha_{HEN}$) and the radius $ASW_{HEN}^{90\%}$. The distribution of selected HEN candidates is in good agreement with the expected distribution of (upward-going) atmospheric neutrinos. A small fraction of the HEN candidates may however be of cosmic origin and this can be determined by the detection of a GW counterpart as discussed in the next section.

\subsection{Searching for gravitational waves associated with high-energy neutrinos}
\label{Xpipeline}

The GW data used in this analysis were collected during LIGO Virgo S5 VSR1. Combined with ANTARES 5L this yields $\sim$103 days of lifetime.
We used the so-called X-pipeline algorithm (~\cite{Sutton:2009gi}) to search for unmodelled GW bursts (duration $\leq 1$ s) in association with each of the selected 216 HEN candidates.  For this analysis the circular polarization of the impinging GW is assumed. The 1000 second duration data segments around the HEN trigger time $t_{HEN}$ defines the ``on-source'' segments. The ``on-source'' data from all available GW detectors are searched coherently over the sky region identified by the $ASW_{GW+HEN}^{90\%}$ (see Sec.~\ref{asw}). The same analysis is applied to two ``off-source'' segments (3 hour duration) surrounding the ``on-source'' region.

The X pipeline estimates the significance of each GW candidate event detected in the ``on-source'' by computing the rate of occurrence of a similar transient in the ``off-source'' segments. The list of final candidate events is subjected to additional checks that may result in vetoing events overlapping in time with known instrumental or environmental disturbances (~\cite{Abbott:2009kk}). 

The first joint GW and HEN analysis is complete and is currently under internal review. The final results will be published in an article in preparation. We could however estimate the search sensitivity for various emission models. As stated earlier, coalescing binaries of neutron stars are of particular interest for this study as they may be associated with the release of GRBs. We estimated the distance reach for those sources to be $\sim$ 8 Mpc (inferred from GW data only). Using typical assumptions on the HEN production in association with GRB, we obtain a similar distance reach (for at least one neutrino observed by ANTARES). We conclude that the distance reach of the joint search is of the same order.

\section{Acknowledgements}

\begin{acknowledgements}
The authors gratefully acknowledge the support of the United States National Science Foundation for the construction and operation of the LIGO Laboratory, the Science and Technology Facilities Council of the United Kingdom, the Max-Planck-Society, and the State of Niedersachsen/Germany for support of the construction and operation of the GEO600 detector, and the Italian Istituto Nazionale di Fisica Nucleare and the French Centre National de la Recherche Scientifique for the construction and operation of the Virgo detector. The authors also gratefully acknowledge the support of the research by these agencies and by the Australian Research Council, the International Science Linkages program of the Commonwealth of Australia, the Council of Scientific and Industrial Research of India, the Istituto Nazionale di Fisica Nucleare of Italy, the Spanish Ministerio de Educaci\'on y Ciencia, the Conselleria d'Economia Hisenda i Innovaci\'o of the Govern de les Illes Balears, the Foundation for Fundamental Research on Matter supported by the Netherlands Organisation for Scientific Research, the Polish Ministry of Science and Higher Education, the FOCUS Programme of Foundation for Polish Science, the Royal Society, the Scottish Funding Council, the Scottish Universities Physics Alliance, The National Aeronautics and Space Administration, the Carnegie Trust, the Leverhulme Trust, the David and Lucile Packard Foundation, the Research Corporation, and the Alfred P. Sloan Foundation. 

The authors also acknowledge the financial support of the funding agencies for the construction and operation of the ANTARES neutrino telescope: Centre National de la Recherche Scientifique (CNRS), Commissariat \`a l\'energie atomique et aux \'energies alternatives (CEA), Agence National de la Recherche (ANR), Commission Europ\'eenne (FEDER fund and Marie Curie Program), R\'egion Alsace (contrat CPER), R\'egion Provence-Alpes-C\^ote d'Azur, D\'epartement du Var and Ville de La Seyne-sur-Mer, France; Bundesministerium fur Bildung und Forschung (BMBF), Germany; Istituto Nazionale di Fisica Nucleare (INFN), Italy; Stichting voor Fundamenteel Onderzoek der Materie (FOM), Nederlandse organisatie voor Wetenschappelijk Onderzoek (NWO), the Netherlands; Council of the President of the Russian Federation for young scientists and leading scientific schools supporting grants, Russia; National Authority for Scientific Research (ANCS), Romania; Ministerio de Ciencia e Innovacion (MICINN), Prometeo of Generalitat Valenciana (GVA) and Multi-Dark, Spain. They also acknowledge the technical support of Ifremer, AIM and Foselev Marine for the sea operation and the CC-IN2P3 for the computing facilities. 

This work has received support from the Groupement de Recherche Ph\'enom\`enes Cosmiques de Haute Energie. This publication has been assigned LIGO Document Number LIGO-P1100127. 
\end{acknowledgements}


\bibliographystyle{aa}  
\bibliography{sf2a-template} 

\begin{thebibliography}{14}
\expandafter\ifx\csname natexlab\endcsname\relax\def\natexlab#1{#1}\fi

\bibitem[{Abadie {et~al.}(2010)}]{Abadie:2010uf}
Abadie, J. {et~al.} 2010, Astrophys. J., 715, 1453

\bibitem[{Abbott {et~al.}(2004)}]{abbott:2004:517}
Abbott, B. {et~al.} 2004, NUCL. INSTRUM. METH. A, 517, 154

\bibitem[{Abbott {et~al.}(2010)}]{Abbott:2009kk}
Abbott, B.~P. {et~al.} 2010, Astrophys. J., 715, 1438

\bibitem[{Acernese {et~al.}(2008)}]{Virgo}
Acernese, F. {et~al.} 2008, Class. Quantum Gravity, 25, 114

\bibitem[{Ageron {et~al.}(2007)}]{Ageron:2007rj}
Ageron, M. {et~al.} 2007, Nucl. Instrum. Meth., A578, 498

\bibitem[{Aguilar {et~al.}(2011{\natexlab{a}})}]{Collaboration:2011nsa}
Aguilar, J. {et~al.} 2011{\natexlab{a}}, Nucl. Instrum. Meth., A656, 11

\bibitem[{Aguilar {et~al.}(2011{\natexlab{b}})}]{Aguilar:2011zz}
Aguilar, J.~A. {et~al.} 2011{\natexlab{b}}, Astropart. Phys., 34, 652

\bibitem[{Aguilar {et~al.}(2011{\natexlab{c}})}]{Aguilar:2010sf}
Aguilar, J.~A. {et~al.} 2011{\natexlab{c}}, Astropart. Phys., 34, 539

\bibitem[{Baret {et~al.}(2011)}]{Baret:2011tk}
Baret, B. {et~al.} 2011, Astropart. Phys., 35, 1

\bibitem[{Becker(2008)}]{Becker:2007sv}
Becker, J.~K. 2008, Phys.Rept., 458, 173

\bibitem[{Chassande-Mottin {et~al.}(2010)}]{eric}
Chassande-Mottin, E. {et~al.} 2010, J. Phys.: Conf. Ser., 243

\bibitem[{Smith(2009)}]{Smith:2009bx}
Smith, J.~R. 2009, Class. Quant. Grav., 26, 114013

\bibitem[{Sutton {et~al.}(2010)}]{Sutton:2009gi}
Sutton, P.~J. {et~al.} 2010, New J. Phys., 12, 053034

\bibitem[{Van~Elewyck {et~al.}(2009)}]{VanElewyck:2009pf}
Van~Elewyck, V. {et~al.} 2009, Int. J. Mod. Phys., D18, 1655

\end{thebibliography}

\end{document}